\documentclass[12pt]{iopart}

\usepackage{iopams}    
\usepackage{setstack}  

\usepackage{graphicx}
\usepackage{color}
\usepackage{hyperref}
\usepackage{framed}
\usepackage{cite}
\usepackage{soul}

\newcommand{\beq}{\begin{equation}}
\newcommand{\eeq}{\end{equation}}
\newcommand{\ba}[1]{\begin{array}{#1}}
\newcommand{\ea}{\end{array}}
\newcommand{\bea}{\begin{eqnarray}}
\newcommand{\eea}{\end{eqnarray}}

\newcommand{\ben}{\begin{enumerate}}
\newcommand{\een}{\end{enumerate}}
\newcommand{\bit}{\begin{itemize}}
\newcommand{\eit}{\end{itemize}}
\newcommand{\bde}{\begin{description}}
\newcommand{\ede}{\end{description}}

\newcommand{\ds}{\displaystyle}

\newcommand{\avg}[1]{\langle {#1} \rangle}

\begin{document}

\title{Effect of shortest path multiplicity on congestion of multiplex networks}

\author{Albert Sol\'e-Ribalta$^{1,2}$, Alex Arenas$^3$, Sergio G\'omez$^3$}

\address{$^1$ Internet Interdisciplinary Institute, Universitat Oberta de Catalunya, 08018 Barcelona, Catalonia, Spain}
\address{$^2$ URPP Social Networks, University of Zurich, CH-8050 Zurich, Switzerland}
\address{$^3$ Departament d'Enginyeria Inform\`atica i Matem\`atiques, Universitat Rovira i Virgili, 43007 Tarragona, Catalonia, Spain}

\eads{\mailto{asolerib@uoc.edu}, \mailto{alexandre.arenas@urv.cat}, \mailto{sergio.gomez@urv.cat}}

\begin{abstract}

Shortest paths are representative of discrete geodesic distances in graphs, and many descriptors of networks depend on their counting. In multiplex networks, this counting is radically important to quantify the switch between layers and it has crucial implications in the transportation efficiency and congestion processes. Here we present a mathematical approach to the computation of the joint distribution of distance and multiplicity (degeneration) of shortest paths in multiplex networks, and exploit its relation to congestion processes. The results allow to approximate semi-analytically the onset of congestion in  multiplex networks as a function of the congestion of its layers.

\end{abstract}

\maketitle

\section{Introduction}

Shortest paths in graphs are defined as paths between two nodes such that the sum of the weights of the links composing the path is minimized \cite{noh2002stability}. The computation of shortest paths is of utmost importance to determine the efficiency of the network to exchange information \cite{latora2001efficiency}, to compute the load of nodes (defined as the number of shortest path traversing it) \cite{newman2001scientific,goh2001uni,holme2002vertex,motter2002cascade}, to predict and alleviate congestion \cite{guimera2002optimal,guimera2002dynamical,sole2016model,sole2018decongestion} or to classify networks \cite{goh2002classification}, to mention some. The main applications of the finding of shortest paths are routing strategies \cite{zhang2007adaptive,yan2006efficient}, analysis of road networks \cite{zhan1998shortest}, epidemic spreading \cite{brockmann2013hidden}, or analysis of brain activity \cite{rubinov2010complex}, among others.

The development of efficient numerical algorithms for both the exact and approximate calculation of shortest paths is a problem by itself that still attracts the attention of computer scientists \cite{Agarwal2018}. The exact numerical calculation of all the shortest paths from a single node is commonly addressed using the well-known Dijkstra's algorithm whose running time is $O(M + N \log N)$, where $N$ is the number of nodes and $M$ is the number of edges of the graph. Given that, the numerical calculation of shortest paths in large networks is still computationally expensive.

In particular, the computation of shortest paths is essential for the determination of a fundamental measure of centrality: edge and node betweenness. The node (or link) betweenness is normally defined as the fraction of shortest paths between node pairs that pass through the node (or link) of interest. The ranking according to betweenness \cite{freeman1977set} is informative about the bottlenecks of a certain network structure, and can be used to determine the most critical node or link with respect to any process traversing the network using shortest paths. In particular, the node with the largest betweenness in a network defines the onset of congestion \cite{guimera2002optimal,zhao2005onset}.

The extension of centrality descriptors to multiplex networks \cite{de2013mathematical,de2014navigability,de2015ranking} prompts for the quantification of the distribution of shortest paths in this new scenario \cite{sole2014centrality,sole2016random}. Multiplex networks are defined as a set of interconnected layers of networks, where every node has its own replica through all layers, and is connected with them \cite{kivela2014multilayer,boccaletti2014structure,de2016physics,bianconi2018multilayer}. Shortest paths in multiplex networks are essentially governed by the shortest paths at each layer, but also by those emerging from partial paths at different layers and the switch between layers. Here, we present a method to analytically determine the distribution of shortest paths in multiplex networks, with special emphasis on the multiplicity of shortest paths occurring in the mentioned set up. This computation allows us to determine semi-analytically the feasible region of congestion induced by the multiplex structure \cite{sole2016congestion}.

The paper is structured as follows. First, we present the definition of shortest paths in multiplex networks. Next, we show how to compute the distribution of shortest paths in two-layer multiplex networks. Once we provide a way to compute the joint distribution of shortest paths and their multiplicities in sigle layer Erd\H{o}s-R\'enyi networks, we extend the method to multiplex network, and evaluate the accuracy of our analytical predictions. Finally, we apply our results to the analysis of congestion in multiplex networks, being able to predict the region of parameters in which the multiplex structure may induce congestion.

\section{Shortest paths in multiplex networks}

\begin{figure}[t]
  \begin{framed}
    \begin{center}
      \begin{tabular}{ll}
        (a) & (b)
        \\
        \includegraphics[width=0.48\textwidth]{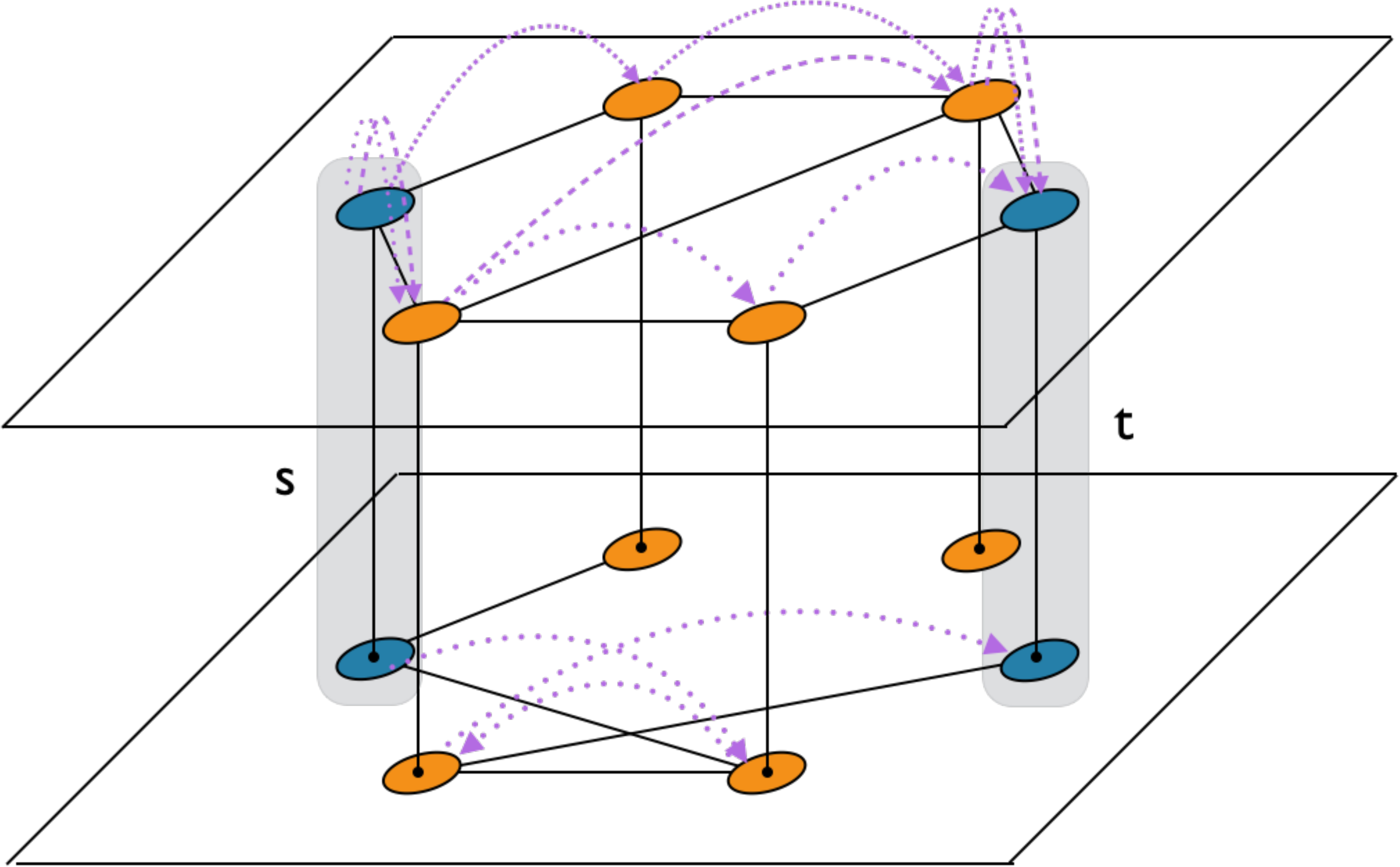}
        &
        \includegraphics[width=0.48\textwidth]{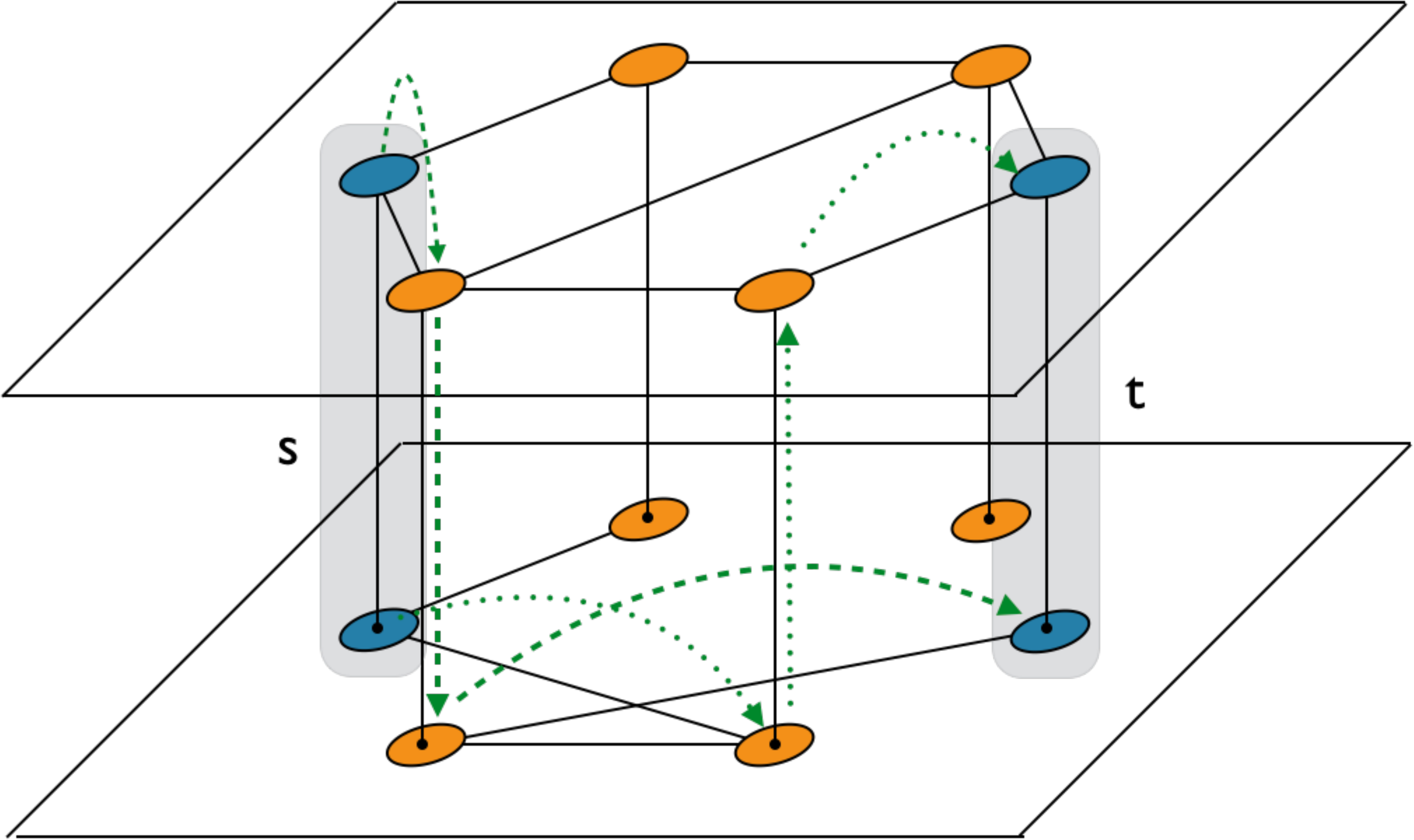}
      \end{tabular}
      \caption{Shortest paths in multiplex networks. An element traveling between locations $s$ and $t$ on the multiplex structure has to choose a path among the possible ones shown in the figure. There are $6$~shortest paths between~$s$ and~$t$, all of them of length~$3$. Purple lines in panel (a) show classical single layer shortest paths, laying within a unique layer. Green lines in panel (b) show multiplex shortest paths, which make use of interlayer links to jump between layers.}
      \label{fig:paths_in_multiplex}
    \end{center}
  \end{framed}
\end{figure}

Multiplex networks may, at first sight, seem equivalent to single layer networks, just with the difference of having two kinds of edges: intralayer links between nodes in the same layer, and interlayer links connecting the replicas of each node in the different layers (see \fref{fig:paths_in_multiplex}). However, the fact that the replicas of each node refer exactly to the same entity, has important consequences on the structural and dynamical properties of multiplex networks.

Suppose, for example, we have the transportation network of a city, formed by metro and bus connections. Nodes represent the locations of the metro stations and bus stops, and interlayer links allow the transfer from metro to bus and vice versa. There exists a cost associated with each link, which accounts for the time needed to travel from one location to another (intralayer links), and the time spent to make a transfer (interlayer links). A path in this network must take into account both types of cost, thus we could use any of the standard methods to find the shortest paths in graphs. Now, consider that we want to go from a source location~$A$ to a destination location~$B$ taking the shortest route, and that we may start the trip either with bus ($A1$) or metro ($A2$). The algorithm provides us with four different types of paths: two in which the origin and destination are both in the same layer, $A1 \rightarrow B1$ and $A2 \rightarrow B2$; and two where the endpoints belong to different layers, $A1 \rightarrow B2$ and $A2 \rightarrow B1$. Note that neither of the paths are necessary contained within a single layer, e.g., a shortest path between $A1$ and $B1$ could change layer twice. Since we are only interested in going from~$A$ to~$B$, and it is irrelevant if we start or end the trip in the bus or metro, some of these shortest paths should be discarded. For example, if the length of the shortest paths $A1 \rightarrow B1$ is larger than that for $A1 \rightarrow B2$, we should discard ending our traversal at layer~1. These means that, in two-layer multiplex networks, shortest paths between two nodes may exist in just one layer, in the other layer, in a path with transfers (what we call \emph{multiplex paths}), or in a combination of them (see \fref{fig:paths_in_multiplex}). Consequently, any property depending on paths in multiplex networks must consider the concept of source and destination as being intrinsically different to that of nodes, such as in the calculation of centrality measures \cite{sole2014centrality,de2015ranking,sole2016random}, the analysis of interdependence \cite{battiston2014}, the analysis of random walks \cite{de2014navigability}, or the study of congestion \cite{sole2016congestion}. The former definition of shortest paths in multiplex networks imposes the necessity of correctly computing their distribution and multiplicity.

\section{Distribution of shortest paths in two-layer multiplex networks}

The problem we are going to address is how the distribution of shortest paths changes when we take two different networks (the layers), and connect them to build a multiplex network. The prescription we propose can be easily extended to more layers, however its combinatorics makes the mathematical analysis more involved.
In particular, we are interested in two important parameters \cite{sole2016congestion}: $\lambda$, the fraction of all shortest paths fully contained in one layer (i.e., paths which do not make use of the multiplex option of changing layer using interlayer links); and $\mu_{\alpha}$, the fraction of non-multiplex shortest paths using only layer~$\alpha$. Thus, $\sum_{\alpha}\mu_{\alpha}=1$, and $1-\lambda$ amounts for the fraction of multiplex shortest paths. It is important to remark that shortest paths are usually \emph{degenerated}, i.e., there are many shortest paths with the same length, with some of them being multiplex paths; we will refer to this degeneration as the \emph{multiplicity} of the shortest paths.

Previous works have provided methods to estimate the complementary cumulative probability $F(d)$ that two randomly selected nodes are separated by a distance larger than~$d$, for certain classes of monoplex (single layer) random networks \cite{fronczak2004average,blondel2007distance,bauckhage2013weibull,steinbock2017distribution,melnik2016simple}. Thus, the probability {distribution $f(d)$ of two nodes being exactly at distance~$d$ can be expressed in terms of the complementary cumulative probability as
\begin{equation}
  f(d) = F(d-1) - F(d)\,.
\end{equation}
We denote by $f_1(d)$, $f_2(d)$ and $f_M(d)$ the respective distributions of shortest paths fully contained in layer~$1$, in layer~$2$, or using the full multiplex structure. Assuming an unweighted and undirected multiplex network with two layers and no degree correlations, we may use this relation to obtain the fraction of multiplex shortest paths:
\begin{eqnarray} \label{eq:lambda}
	1-\lambda = \sum_{d=3}^{\infty} f_M(d)
	& \left[                           F_1(d) F_2(d) \right. \nonumber \\
	& \mbox{} +        \theta_{M1}(d)  f_1(d) F_2(d) \nonumber \\
	& \mbox{} +        \theta_{M2}(d)  F_1(d) f_2(d) \nonumber \\
	& \mbox{} + \left. \theta_{M12}(d) f_1(d) f_2(d) \right]\,.
\end{eqnarray}
The sum starts with $d=3$, since the minimum length of a multiplex path is 3, i.e., at least one hop in the first layer, another in the second layer, and one to perform the change of layer. \Eref{eq:lambda} expresses the four possible ways in which shortest paths can appear in the multiplex: only multiplex paths; multiplex paths mixed with paths in layer~$1$; multiplex paths mixed with paths in layer~$2$; and multiplex paths mixed with paths in layers~$1$ and~$2$. Consider for example the case of multiplex shortest paths mixed with shortest paths in layer~$1$. This correspond to the paths that have  exactly distance $d$ in layer~$1$, $f_1(d)$, and in the multiplex, $f_M(d)$, and distance larger than $d$ in layer~$2$, $F_2(d)$. {For the terms with mixed contributions, the $\theta(d)$ factors capture the fraction of shortest paths that correspond to multiplex paths.} This means we cannot calculate $\lambda$ just with the knowledge of the shortest paths distributions $f(d)$ and $F(d)$, we also need to estimate the multiplicity of the paths of each type.


Let us denote by $P_1(d,\phi)$, $P_2(d,\phi)$ and $P_M(d,\phi)$ the probabilities that randomly chosen source and destination (discarding layer, i.e., locations in our example) are exactly at distance~$d$ with multiplicity~$\phi$, for paths in first layer, second layer, and in the multiplex, respectively. The average multiplicities $\avg{\phi}_1(d)$, $\avg{\phi}_2(d)$ and $\avg{\phi}_M(d)$ for each kind of path can be expressed as:
\numparts
\begin{eqnarray}
  \avg{\phi}_{\alpha}(d) & = & \frac{\ds \sum_{\phi=1}^{\infty} \phi\, P_{\alpha}(d,\phi)}
                                    {\ds \sum_{\phi=1}^{\infty} P_{\alpha}(d,\phi)}\,,
                                    \quad\quad\quad \alpha \in \{1,2\}\,,
  \label{eq:avg_phi_alpha}
  \\
  \avg{\phi}_{M}(d)      & = & \frac{\ds \sum_{\phi=1}^{\infty} \phi\, P_{M}(d,\phi)}
                                    {\ds \sum_{\phi=1}^{\infty} P_{M}(d,\phi)}\,.
  \label{eq:avg_phi_M}
\end{eqnarray}
\endnumparts
Therefore,
\numparts
\begin{eqnarray}
  \theta_{M\alpha}(d) & = & \frac{\ds \avg{\phi}_M(d)}
                                 {\ds \avg{\phi}_M(d) + \avg{\phi}_{\alpha}(d)}\,,
                                 \quad\quad\quad \alpha \in \{1,2\}\,,
  \label{eq:theta_Malpha}
  \\
  \theta_{M12}(d)     & = & \frac{\ds \avg{\phi}_M(d)}
                                 {\ds \avg{\phi}_M(d) + \avg{\phi}_1(d) + \avg{\phi}_2(d)}\,.
  \label{eq:theta_M12}
\end{eqnarray}
\endnumparts
Additionally, the distributions $f(d)$ can be recovered as marginals of $P(d,\phi)$:
\numparts
\begin{eqnarray}
  f_{\alpha}(d) & = & \sum_{\phi=1}^{\infty} P_{\alpha}(d,\phi)\,,
                      \quad\quad\quad \alpha \in \{1,2\}\,,
  \label{eq:f_from_P_Malpha}
  \\
  f_{M}(d) & = & \sum_{\phi=1}^{\infty} P_{M}(d,\phi)\,.
  \label{eq:f_from_P_M12}
\end{eqnarray}
\endnumparts

Similarly to \eref{eq:lambda}, we may use the distributions $F(d)$ and $f(d)$, and the multiplicity factors $\theta(d)$, to calculate how the non-multiplex shortest paths are distributed among the layers. The expression for the first layer reads:
\begin{equation} \label{eq:mu}
  \mu_1 = \sum_{d=1}^{\infty} f_1(d) \left[ F_2(d) + \theta_1(d) f_2(d) \right]\,,
\end{equation}
where
\begin{equation}
  \theta_1(d) = \frac{\ds \avg{\phi}_1(d)}{\ds \avg{\phi}_1(d) + \avg{\phi}_2(d)}\,.
\end{equation}
Equivalent expressions hold for $\mu_2$, and $\mu_1+\mu_2=1$.

Summarizing, we have set all the necessary ingredients to analyze the use of the multiplex structure and of the layers in terms of the joint distributions $P_1(d,\phi)$, $P_2(d,\phi)$ and $P_M(d,\phi)$. In the next sections we develop expressions for each of these probabilities.

\section{Multiplicity of shortest paths in Erd\H{o}s-R\'enyi networks}

The analytical calculation of the joint probability that two nodes are exactly at distance $d$ with multiplicity $\phi$, $P(d,\phi)$, for any type of monoplex networks, is quite involved, thus we are going to restrict our analysis to Erd\H{o}s-R\'enyi networks. However, it would be possible to try with other types of uncorrelated random networks making use of hidden variables as in \cite{fronczak2004average}.

An Erd\H{o}s-R\'enyi network is characterized by two parameters, the number of nodes $N$, and the probability $p$ that an edges exists between any pair of nodes. Thus, we will use the notation $P_{p,N}(d,\phi)$ to refer to the joint distance and multiplicity distribution for these networks. Clearly, at $d=1$,
\numparts
\begin{eqnarray}
  P_{p,N}(1,1) & = & p\,,
  \\
  P_{p,N}(1,\phi) & = & 0\,,\quad \forall\phi>1\,,
\end{eqnarray}
\endnumparts
since we do not allow multiple edges between pairs of nodes. The case $d=2$ can be obtained as the probability of not having a shortest path of length~$1$ times the probability of having exactly $\phi$ different shortest paths of length~$2$. Since paths of length~$2$ have just one intermediate node, we have to choose $\phi$ nodes among the $N-2$ available to build the $\phi$ shortest paths. Formally,
\begin{equation} \label{eq:prob_multiplicity_monoplex_d2}
	P_{p,N}(2,\phi) = (1-p) {N-2 \choose \phi} p^{2\phi}(1-p^2)^{(N-2)-\phi}\,.
\end{equation}
Note that the existence of paths of length~$2$ has probability $p^2$, thus $p^{2\phi}$ expresses the probability of having $\phi$ paths of length~$2$, and $(1-p^2)^{(N-2)-\phi}$ the probability that the rest of the $(N-2)-\phi$ nodes are not used to build shortest paths of length~$2$. Using the same procedure, the generalization to larger distances becomes
\begin{equation} \label{eq:prob_multiplicity_monoplex}
	P_{p,N}(d,\phi) = \left[(1-p)\prod_{k=2}^{d-1}(1-p^k)^{\eta(k)}\right] B(\phi; \eta(d), p^d)\,,
\end{equation}
where
\begin{equation}
  \eta(k)=\prod_{i=0}^{k-2} (N-2-i)
\end{equation}
stands for the amount of possible paths of length~$k\geqslant 2$, and
\begin{equation}
  B(\phi; \eta(d), p^d)={\eta(d) \choose \phi}p^{d\phi}(1-p^d)^{\eta(d)-\phi}
\end{equation}
is the probability density function of the binomial distribution with probability $p^d$. The term within square brackets in \eref{eq:prob_multiplicity_monoplex} computes the probability of not having any path at distance lower than~$d$. Although \eref{eq:prob_multiplicity_monoplex_d2} is exact, \eref{eq:prob_multiplicity_monoplex} is an approximation because it does not consider that some shortest paths may share some (but not all) of their links.

\section{Multiplicity in multiplex networks with Erd\H{o}s-R\'enyi layers}

We now address the computation of the joint distribution $P_M(d,\phi)$ for a multiplex network composed of two Erd\H{o}s-R\'enyi layers, with $N$~nodes each, edge probabilities~$p_1$ and~$p_2$, and without interlayer degree correlations; we will refer to it as $P_{p_1,p_2,N}(d,\phi)$, to emphasize the dependence on these structural parameters. Multiplex shortest paths are characterized by the number of interlayer links they contain, i.e., how many times the path changes from one layer to the other, see \fref{fig:multiplex_paths}. We restrict our analysis to shortest paths that change layer only once, since the number of shortest paths with multiple jumps between layers is usually very small, thus making their contribution negligible, and the generalization of the mathematical formulation for two or more jumps is not difficult but generates very long expressions.

\begin{figure}[t]
  \begin{framed}
    \begin{center}
      \begin{tabular}{l}
        (a)
        \\
        \includegraphics[width=0.65\textwidth]{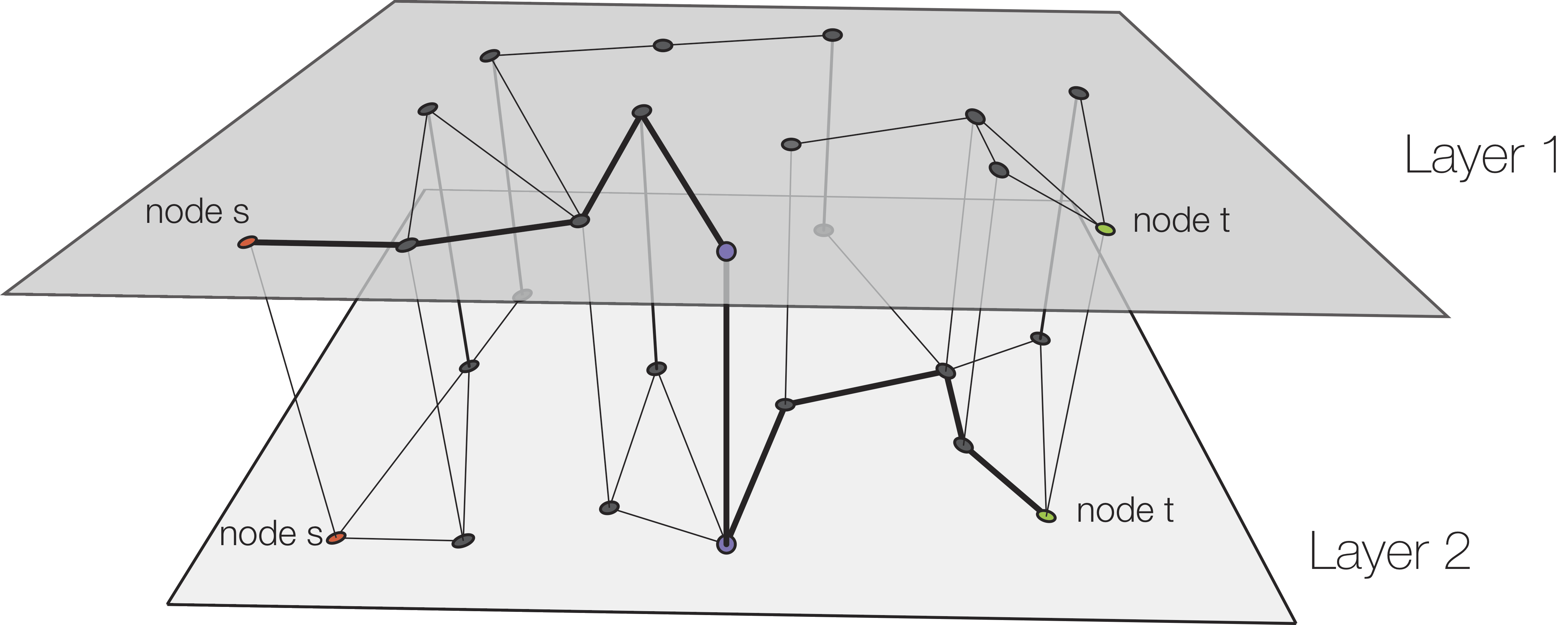}
        \\
        (b)
        \\
        \includegraphics[width=0.65\textwidth]{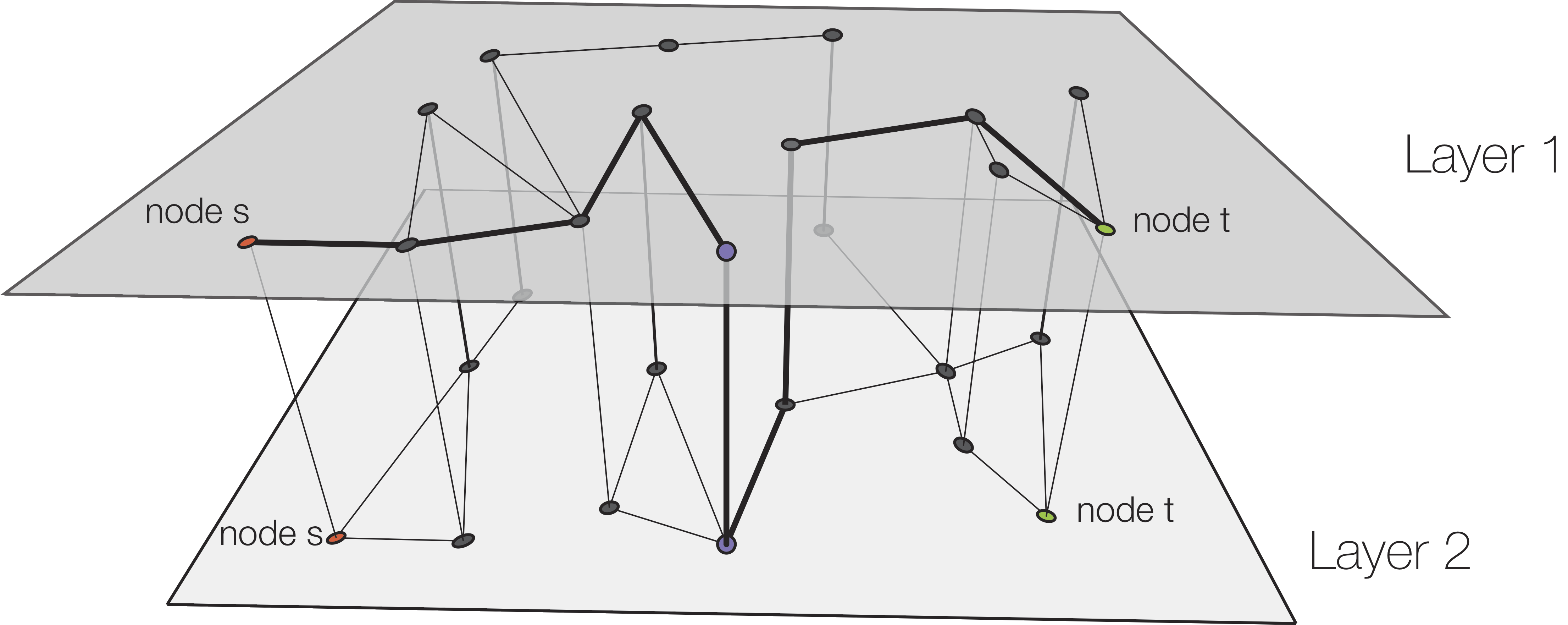}
      \end{tabular}
      \caption{(a) Shortest path in the multiplex network containing a single interlayer link. (b) Shortest path in the multiplex network including two interlayer links.}
      \label{fig:multiplex_paths}
    \end{center}
  \end{framed}
\end{figure}

For each multiplex shortest path which has just one change of layer, as in \fref{fig:multiplex_paths}a, once we have selected the origin and destination, we may choose the intermediate switch node among the remaining $N-2$. Since we are considering a multiplicity $\phi$ of multiplex shortest paths between these origin and destination nodes, the paths may be distributed in many different ways. For example, if $\phi=10$, we could select~$7$ shortest paths to use one intermediate switch node, and the remaining~$3$ to choose another one. The set of all structurally different distributions of one-jump multiplex shortest paths is given by the partition set of the given integer multiplicity $\phi$. The finding and counting of the number of partitions of an integer number constitutes a classical problem in number theory \cite{andrews2004integer}, e.g., they can be enumerated using Young diagrams \cite{young1900quantitative}. We may express the set of available partitions as:
\begin{equation} \label{eq:phi_partitions}
  \fl
	\mathcal{F}(\phi) = \{ \mathbf{\Phi} \in \bigcup_{r=1}^{\phi} \{1,\ldots,\phi\}^r:
	                    \sum_{i=1}^{|\mathbf{\Phi}|} \phi_i = \phi\,,\
	                    \phi_1 \geqslant \cdots \geqslant \phi_{|\mathbf{\Phi}|}\,,\
	                     |\mathbf{\Phi}| \leqslant N-2  \} \,.
\end{equation}
For example, $\phi=4$ can be partitioned in five different ways: $\mathcal{F}(4)=\{(4)$, $(3,1)$, $(2,2)$, $(2,1,1)$, $(1,1,1,1)\}$. If $N=5$, we would have only $3$~possible intermediate nodes, and partition $(1,1,1,1)$ should be discarded, hence the $|\mathbf{\Phi}| \leqslant N-2$ condition added to \eref{eq:phi_partitions}.

Once a partition $\mathbf{\Phi}$ is selected, we need to consider the different ways in which we can choose the intermediate nodes, and the different ways in which the multiplicities $\phi_i$ are assigned to the selected set of intermediate nodes. Gathering together all these contributions, we may write the joint distribution of multiplex shortest paths as:
\begin{equation} \label{eq:prob_multiplicity_multiplex}
  \fl
  P_{p_1,p_2,N}(d,\phi) = \sum_{\mathbf{\Phi} \in \mathcal{F}(\phi)}
                          {N-2 \choose {|\mathbf{\Phi}|}} C(\mathbf{\Phi})
                          \left[
                            R_{p_1,p_2,N}(d)^{N-2-|\mathbf{\Phi}|}
                            \prod_{i=1}^{|\mathbf{\Phi}|} Q_{p_1,p_2,N}(d,{\Phi}_i)
                          \right] \,,
\end{equation}
where the combinatorial number accounts for the selection of intermediate switch nodes, $C(\mathbf{\Phi})$ for the assignment of individual multiplicities, $R_{p_1,p_2,N}(d)$ stands for the probability of not having multiplex paths of length lower or equal than~$d$, and $Q_{p_1,p_2,N}(d,{\Phi}_i)$ stands for the joint probability of having one-switch multiplex shortest paths with known intermediate jump node. Factor $C(\mathbf{\Phi})$ is just the multinomial coefficient corresponding to the frequencies of the components of $\mathbf{\Phi}$. For example, if $\mathbf{\Phi}=(14,3,3)$, there are three components, the~$14$ with frequency~$1$ and the~$3$ with frequency~$2$, leading to $C(\mathbf{\Phi})=3!/(1!\,2!)=3$, which corresponds to the $3$~ways in which we can sort the components of $\mathbf{\Phi}$, i.e., $(14,3,3)$, $(3,14,3)$ and $(3,3,14)$.

Probability $R_{p_1,p_2,N}(d)$ admits a simple expression:
\begin{equation} \label{eq:prob_r}
  R_{p_1,p_2,N}(d) = 1 - \sum_{s=1}^d \sum_{\phi=1}^{\infty} Q_{p_1,p_2,N}(s,\phi)\,,
\end{equation}
However, $Q_{p_1,p_2,N}(d,\phi)$ still requires further decompositions. Although we know the origen, destination and intermediate nodes, the total length~$d$ of the paths, and the total multiplicity~$\phi$, it remains to establish: the layer~$\alpha$ at which the path starts; the length~$\ell$ covered in layer~$1$ (the rest of the path in layer~$2$ will have length $d-\ell-1$); the distribution of the multiplicities between the two layers. Moreover, when~$\phi>1$, we may have a combination of paths starting at different layers, with different lengths in each layer, and with different distribution of the multiplicities per layer. Denoting by $\psi_{\alpha,\ell,\beta}$ the multiplicity of paths in layer~$\beta$, when they start in layer~$\alpha$ and have length~$\ell$ in layer~$1$, the set of possible decompositions of the multiplicity can be written as:
\begin{equation} \label{eq:phi_decomposition}
	\mathcal{G}(d,\phi) = \{ \mathbf{\Psi} \in \{0,\ldots,\phi\}^{2\times(d-2)\times 2}:
	                      \sum_{\alpha=1}^{2} \sum_{\ell=1}^{d-2}
	                      \psi_{\alpha,\ell,1} \psi_{\alpha,\ell,2} = \phi \}\,.
\end{equation}
Note that $a$~paths arriving to the switching node in one layer, and $b$~paths departing from it in the other layer, generate $a b$~different paths in the multiplex, hence the products inside the sums in \eref{eq:phi_decomposition}. An example of a decomposition $\mathbf{\Psi}$ of $\mathcal{G}(5,10)$ could be: $2$~paths of length~$2$ starting in layer~$1$, followed by $3$~paths of length~$2$ in layer~$2$ (amounting $6$~paths), plus $1$~path of length~$1$ starting in layer~$2$, followed by $4$~paths of length~$3$ in layer~$1$ (amounting $4$~paths), which correspond to $\psi_{1,2,1}=2$, $\psi_{1,2,2}=3$, $\psi_{2,3,2}=1$ and $\psi_{2,3,1}=4$ (the rest of the components are zero). In practical terms, the calculation of $\mathcal{G}(d,\phi)$ involves: finding the partitions of $\phi$; factorizing the parts as products of two integers, by calculating all their divisors; choosing initial layer; and dividing the total length as the sum of the lengths in each layer plus one.

Using the decomposition set in \eref{eq:phi_decomposition}, we may finally calculate the remaining probabilities in \eref{eq:prob_multiplicity_multiplex} and \eref{eq:prob_r}:
\begin{equation} \label{eq:prob_q}
  \fl
  Q_{p_1,p_2,N}(d,\phi) = \sum_{\mathbf{\Psi} \in \mathcal{G}(d,\phi)}
                          \underset{\psi_{\alpha,\ell,1}\psi_{\alpha,\ell,2}\neq 0}
                                   {\prod_{\alpha=1}^{2} \prod_{\ell=1}^{d-2}}
                          P_{p_1,N}(\ell,\psi_{\alpha,\ell,1})
                          P_{p_2,N}(d-\ell-1,\psi_{\alpha,\ell,2})\,.
\end{equation}
Note the presence of the joint probabilities \eref{eq:prob_multiplicity_monoplex} for single layer Erd\H{o}s-R\'enyi networks, which account for the subpaths of the multiplex shortest paths fully contained in each of the layers.

\section{Evaluation of analytical predictions}

To check the validity of our calculations of the distribution of paths in multiplex networks, we have generated a large set of two-layer multiplex networks with uncorrelated Erd\H{o}s-R\'enyi layers, and compared the predicted values with the experimental ones. In particular, we have analyzed multiplex networks with 500~nodes in each layer, for a total of $50^2$ different configurations. These configurations correspond to $50$~logarithmically spaced values of the average degree $\avg{k}$ of each layer, with values ranging between~$5$ and~$35$. For each configuration, we generate 100~different multiplex networks, calculate their experimental values of $\lambda$ and $\mu_1$, and take averages. These averages are then compared with the predicted values using \eref{eq:lambda} and \eref{eq:mu}, respectively. The analytical values only consider multiplex shortest paths with just one change of layer (as discussed above), and contributions from multiplicities larger than~100 have been discarded. The results are presented in \fref{fig:teo_vs_mc}, which shows an excellent agreement between theory and experiments.

\begin{figure}[t]
  \begin{framed}
    \begin{center}
      \begin{tabular}{ll}
        (a) & (b)
        \\
        \includegraphics[width=0.49\textwidth]{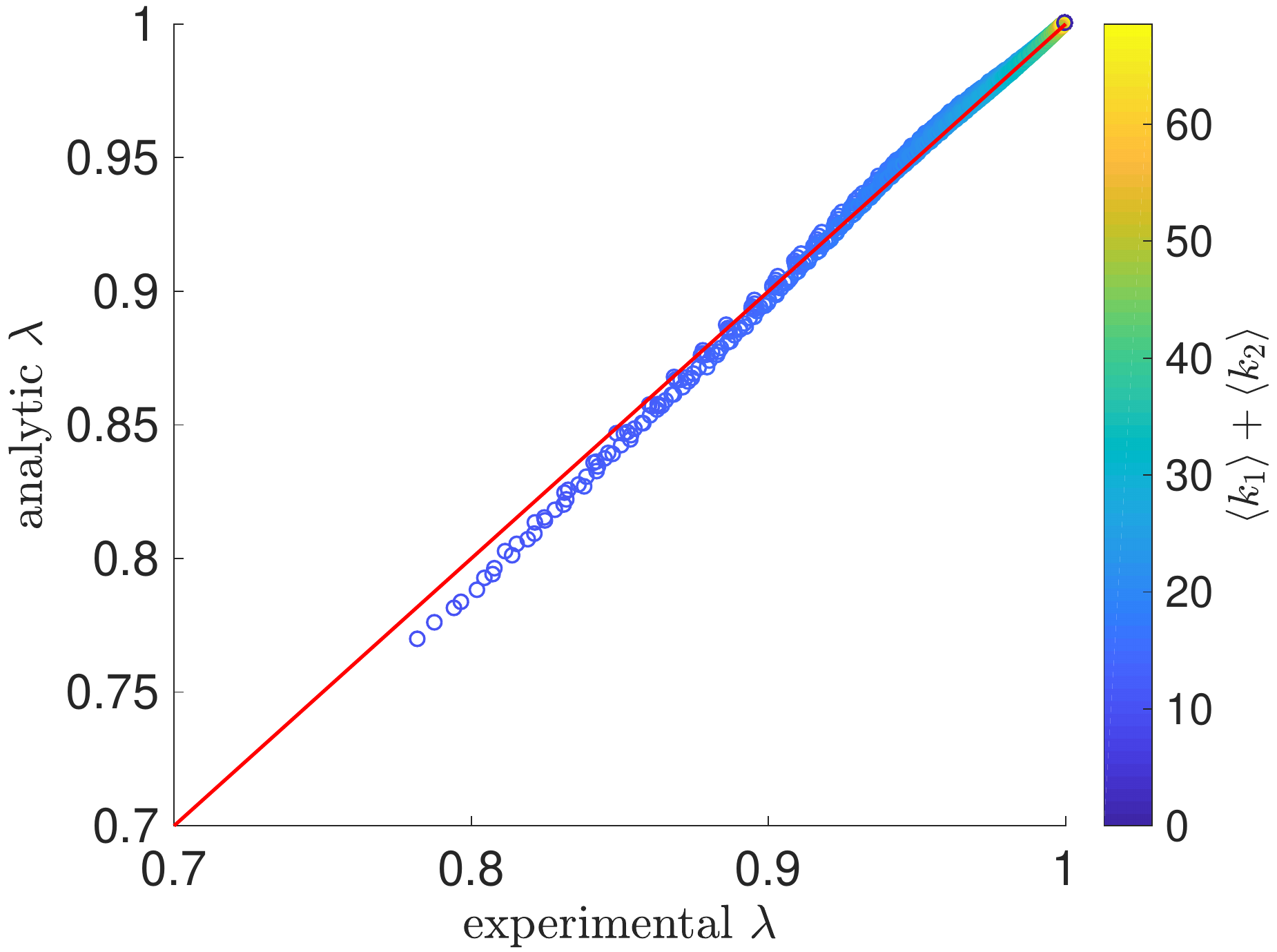}
        &
        \includegraphics[width=0.49\textwidth]{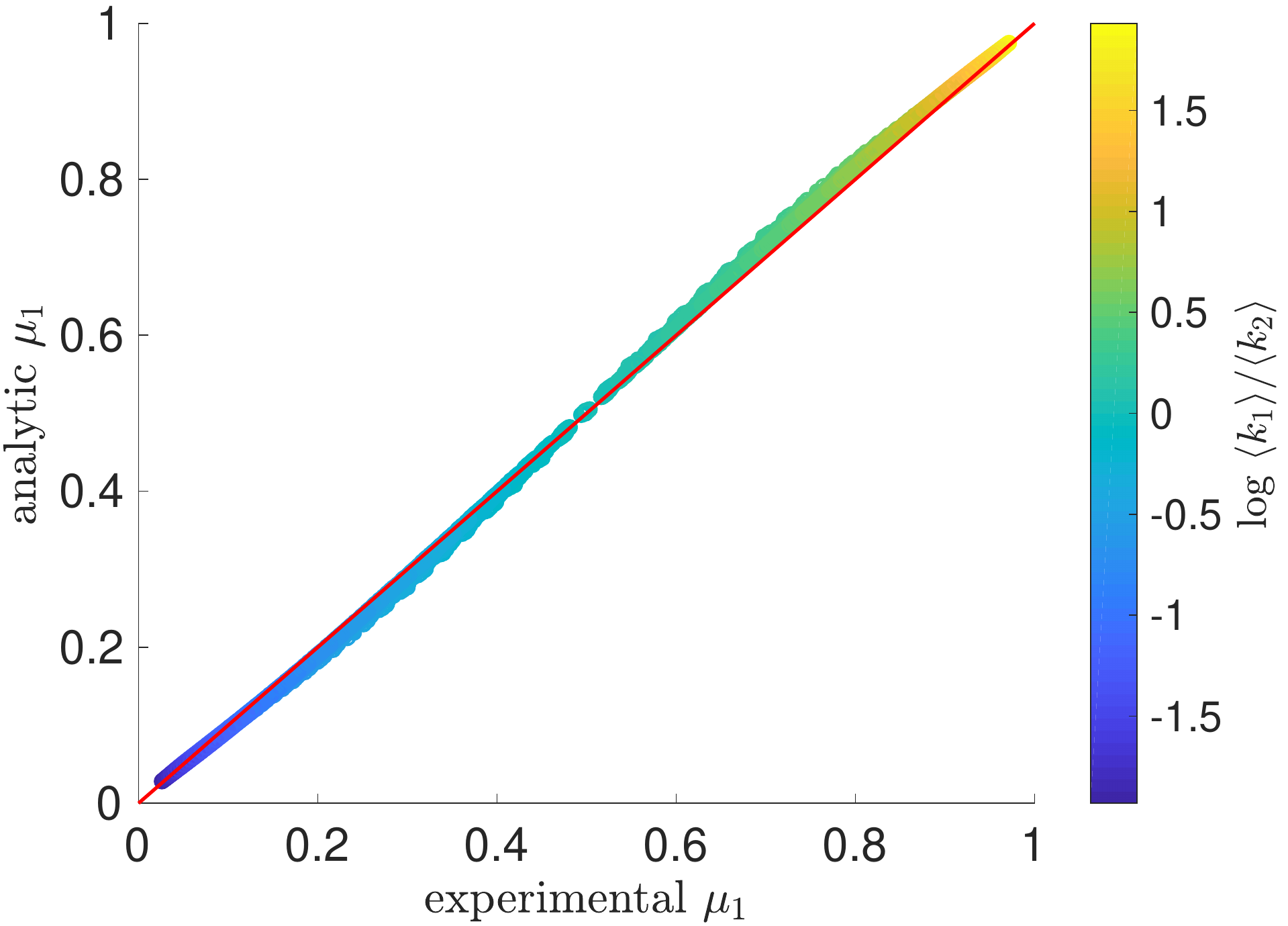}
      \end{tabular}
      \caption{(a) Evaluation of the accuracy of \eref{eq:lambda} on predicting~$\lambda$, the fraction of non-multiplex shortest paths. (b) Evaluation of the accuracy of \eref{eq:mu} on predicting~$\mu_1$, the fraction of non-multiplex shortest paths contained in the first layer. Each experimental point is the average of $100$~two-layer multiplex networks with Erd{\H{o}}s-R\'{e}nyi layers.}
      \label{fig:teo_vs_mc}
    \end{center}
  \end{framed}
\end{figure}

When the average degree of an Erd\H{o}s-R\'enyi network is large, its average shortest path length is small \cite{fronczak2004average}, thus it is difficult to find multiplex shortest paths since the overhead of changing layer has to be compensated with very short paths in the layers. As a consequence, the fraction~$\lambda$ of shortest paths fully included in one of the layers tends to~$1$, as shown in \fref{fig:teo_vs_mc}a. If we reduce these average degrees, the shortest paths of the layers become larger, and the opportunities to find multiplex shortest paths increases, yielding lower values of~$\lambda$. In the extreme cases in which the average degrees are very small, multiplex shortest paths become more common, and it is even possible to find shortest paths with two or more changes of layer; this is the reason for the small deviations between the experimental and predicted values of~$\lambda$ for small $\avg{k_1}+\avg{k_2}$.

\Fref{fig:teo_vs_mc}b shows also that non-multiplex shortest paths tend to be concentrated in the layer with largest average degree, i.e., with smaller average shortest path length. Thus, parameter~$\mu_1$ approaches value~$1$ when $\avg{k_1}\gg\avg{k_2}$, and~$0$ when $\avg{k_1}\ll\avg{k_2}$.

\section{Congestion in multiplex networks}

Shortest paths play an important role in congestion phenomena in complex networks. When elements (packages, vehicles, etc.) travel a network using shortest paths, the load of the nodes is directly related to the number of shortest paths that make use of them. If the capacity of some nodes to process these elements is lower than their incoming rate, congestion emerges \cite{guimera2002optimal}. In the general case of multiplex networks \cite{sole2016congestion}, it was shown that congestion appears when the injection rate per node, $\rho$, reaches a critical value $\rho_c$:
\begin{equation} \label{eq:rho_c}
  \rho_c = \tau L^{-1} \frac{N-1}{\mathcal{B}^{\ast}}\,,
\end{equation}
where $\tau$ is the processing capacity of the nodes, and $\mathcal{B}^{\ast}$ is the maximum betweenness of all nodes in all layers of the multiplex network. Additionaly, as it is shown in \cite{sole2016congestion}, the maximum betweenness found in the multiplex can be approximated in terms of the betweenness of the most efficient layer, $\ell$, as $\mathcal{B}^{\ast} \approx \lambda \mu_{\ell} \mathcal{B}_{\ell}^{\ast}\,$. Specifically, $\ell$ is the layer for which the onset of congestion is larger than for the rest of the layers, when layers are considered as independent networks. Thus, the onset of congestion $\rho_c$ becomes
\begin{equation} \label{eq:rho_c_approx}
  \rho_c \approx \frac{1}{L\lambda\mu_{\ell}} \rho_c^{(\ell)}\,,
\end{equation}
where $\rho_c^{(\ell)}$ is the critical injection rate of layer $\ell$, given by (see \cite{guimera2002optimal})
\begin{equation} \label{eq:rho_c_approx_2}
  \rho_c^{(\ell)} = \tau \frac{N-1}{\mathcal{B}^{\ast}_{\ell}}\,,
\end{equation}
For example, if layer one is the most efficient, $\ell=1$, and we have two layers, $L=2$, the first layer of the multiplex accepts much more load than the second layer before congestion appears, thus $\rho_c^{(1)}>\rho_c^{(2)}$. When we connect these two layers to form a multiplex, there exists a migration of shortest paths from the second to the first layer, which increases its load, eventually becoming responsible for the onset of congestion of the multiplex. This increases the congestion of the most efficient layer, as shown in \eref{eq:rho_c_approx}. An important consequence of this redistribution of shortest paths in the multiplex, and the appearance of multiplex shortest paths, is that the multiplex may attain congestion with lower load than for any of its separated layers, the so-called congestion induced by the multiplex structure \cite{sole2016congestion}.
In our example, this happens when $\rho_c < \rho_c^{(2)}$. By combining this inequality with \eref{eq:rho_c_approx} and \eref{eq:rho_c_approx_2}, we obtain the condition for having congestion induced by the multiplex:
\begin{equation} \label{eq:congestion_induced}
	L\lambda\mu_{1}\leqslant \frac{\mathcal{B}^{\ast}_2}{\mathcal{B}^{\ast}_1}\,.
\end{equation}

We have shown above how to analytically estimate the values of $\lambda$ and $\mu_1$, by using \eref{eq:lambda} and \eref{eq:mu}, respectively. However, the estimation of $\mathcal{B}^{\ast}_{\alpha}$ in terms of the structural properties of the network is not straightforward. We have experimentally observed that the relationship between the maximum betweenness (excluding endpoints) and $\avg{k_{\alpha}}$ follows a power law with exponent $-1$, with a constant proportional to $N-1$ and to the average multiplicity found in Erd\H{o}s-R\'enyi networks of the same size, $\avg{\phi}^{\ast}$. Thus, we may express the maximum betweenness as
\begin{equation} \label{eq:betweenness}
  \mathcal{B}^{\ast}_\alpha \approx (N-1) \avg{\phi}^{\ast}\avg{k_{\alpha}}^{-1} + 2(N-1)\,,
\end{equation}
where
\begin{equation}
  \avg{\phi}^{\ast} = \max_{2\leq \avg{k} \leq N}
    \frac{1}{d_{\avg{k}}} \sum_{d=1}^{d_{\avg{k}}} \avg{\phi}_{\avg{k}}(d)\,.
\end{equation}
The term $d_{\avg{k}}$ corresponds to the maximum expected distance in an Erd\H{o}s-R\'enyi network with average degree $\avg{k}$, and $\avg{\phi}_{\avg{k}}(d)$ to the average multiplicity in \eref{eq:avg_phi_alpha} for that network.

\begin{figure}[t]
  \begin{framed}
    \begin{center}
      \includegraphics[width=0.65\textwidth]{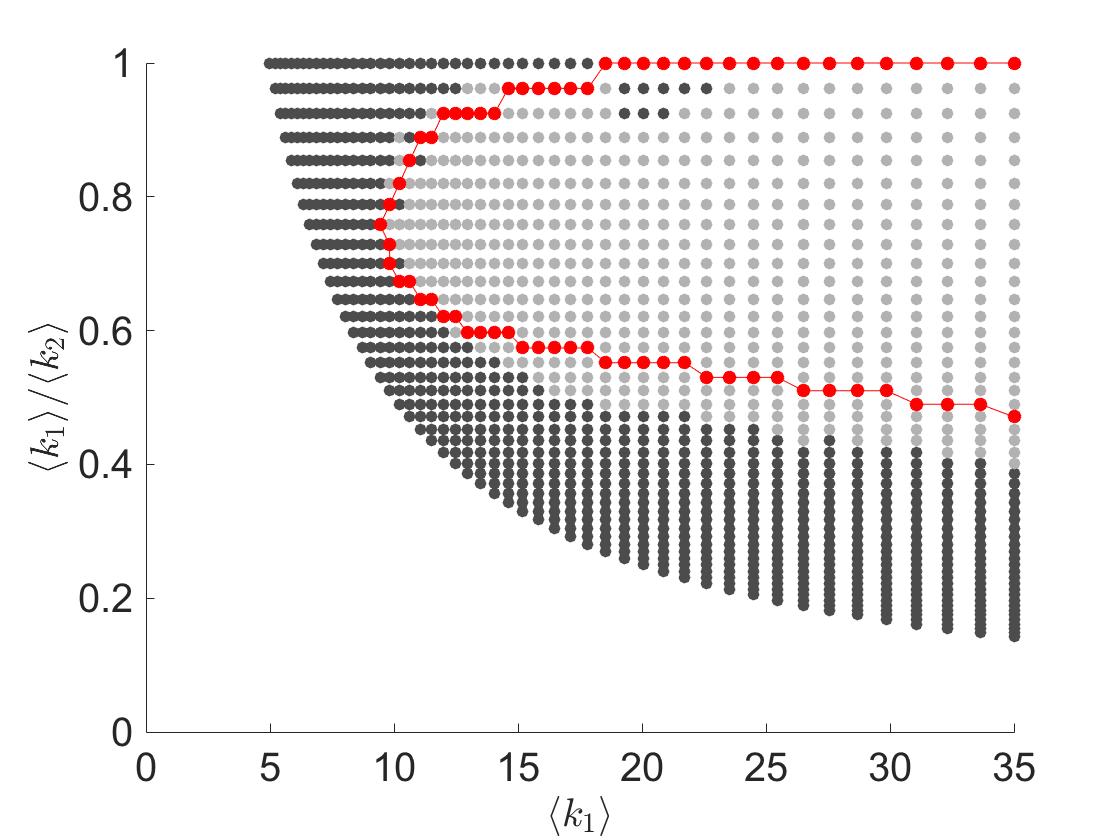}
      \caption{Estimation of the congestion induced by the multiplex structure. Light grey region corresponds to configurations where the multiplex induces congestion, and dark grey to regions where the multiplex structure does not induce congestion. Red line represents our semi-analytical estimation of the frontier between these two regions, using \eref{eq:congestion_induced} and \eref{eq:betweenness}.}
      \label{fig:congestion}
    \end{center}
  \end{framed}
\end{figure}

We show in \fref{fig:congestion} the estimated region of the parameters space in which congestion is predicted to be induced by the multiplex structure, for multiplex networks composed of Erd\H{o}s-R\'enyi layers, in good agreement with the experimental results.

\section{Conclusions}

We have presented a method to compute the multiplicity and distribution of shortest paths in multiplex networks. Using the method, we have analytically determined the distribution and multiplicity of shortest paths in duplex of Erd\H{o}s-R\'enyi networks. This results are essential to determine the onset of congestion on multiplex structures, as for example in many multimodal transportation networks. We can use the analytical findings to determine the area of the phase diagram where the multiplex can induce congestion. This work is relevant for any dynamical process that uses shortest path as a routing strategy in multilayer networks.

\section*{Acknowledgements}

This work has been supported by Ministerio de Econom\'{\i}a y Competitividad (grant FIS2015-71582-C2-1), Generalitat de Catalunya (grant 2017SGR-896), and Universitat Rovira i Virgili (grant 2017PFR-URV-B2-41). AA acknowledges partial financial support from ICREA Academia and the James S.\ McDonnell Foundation.

\section*{ORCID iDs}

\noindent
Albert Sol\'e-Ribalta \url{https://orcid.org/0000-0002-2953-5338} \\
Alex Arenas         \url{https://orcid.org/0000-0003-0937-0334} \\
Sergio G\'omez        \url{https://orcid.org/0000-0003-1820-0062}

\section*{References}


\providecommand{\newblock}{}

\end{document}